\documentclass[twocolumn]{aastex63}

\usepackage{amsmath}
\usepackage{csquotes} 

\usepackage{rotating}

\newcommand{\degree}{\ensuremath{^\circ}}

\def \sun {$_{\odot}$}

\received{April 27, 2020}
\revised{August 5, 2020}
\accepted{\today}
\submitjournal{ApJ}

\shorttitle{The Magnetic Field of L183}
\shortauthors{Karoly et al.}

\begin{document}

\title{Revisiting the Magnetic Field of the L183 Starless Core}

\correspondingauthor{Janik Karoly, Archana Soam}
\email{jkaroly@alumni.scu.edu, asoam@usra.edu}


\author{Janik Karoly}
\affiliation{Department of Physics, Santa Clara University, Santa Clara, 500 El Camino Real, Santa Clara, CA, USA} 
\affiliation{SOFIA Science Center, Universities Space Research Association, NASA Ames Research Center, M.S. N232-12, Moffett Field, CA 94035, USA}
\author[0000-0002-6386-2906]{Archana Soam}
\affiliation{SOFIA Science Center, Universities Space Research Association, NASA Ames Research Center, M.S. N232-12, Moffett Field, CA 94035, USA}
\author[0000-0001-6717-0686]{B-G Andersson}
\affiliation{SOFIA Science Center, Universities Space Research Association, NASA Ames Research Center, M.S. N232-12, Moffett Field, CA 94035, USA}
\author[0000-0002-0859-0805]{Simon Coud\'e}
\affiliation{SOFIA Science Center, Universities Space Research Association, NASA Ames Research Center, M.S. N232-12, Moffett Field, CA 94035, USA}
\author[0000-0002-0794-3859]{Pierre Bastien}
\affiliation{Institut de Recherche sur les Exoplan\`etes (iREx) \& Centre de Recherche en Astrophysique du Qu\'ebec (CRAQ),\\ Universit\'e de Montr\'eal, D\'epartement de Physique, C.P. 6128 Succ. Centre-ville, Montr\'eal, QC, H3C 3J7, Canada}
\author{John E. Vaillancourt}
\affiliation{Lincoln Laboratory, Massachusetts Institute of Technology, 244 Wood St., Lexington, MA 02420-9108, USA}
\author{Chang Won Lee}
\affiliation{Korea Astronomy \& Space Science Institute, 776 Daedeokdae-ro, Yuseong-gu, Daejeon, Republic of Korea}
\affiliation{University of Science and Technology, Korea, 217 Gajeong-ro, Yuseong-gu, Daejeon 34113, Republic of Korea}

\begin{abstract}

We present observations of linear polarization from dust thermal emission at 850~$\micron$ towards the starless cloud L183. These data were obtained at the James Clerk Maxwell Telescope (JCMT) using the Submillimetre Common-User Bolometer Array~2 (SCUBA-2) camera in conjunction with its polarimeter POL-2. Polarized dust emission traces the plane-of-sky magnetic field structure in the cloud, thus allowing us to investigate the role of magnetic fields in the formation and evolution of its starless core. To interpret these measurements, we first calculate the dust temperature and column density in L183 by fitting the spectral energy distribution obtained by combining data from the JCMT and the \textit{Herschel} space observatory. We used the Davis-Chandrasekhar-Fermi technique to measure the magnetic field strength in five sub-regions of the cloud, and we find values ranging from $\sim120\pm18~\mu G$ to $\sim270\pm64~\mu G$ in agreement with previous studies. Combined with an average hydrogen column density ($N_{\text{H}_2}$) of $\sim 1.5 \times 10^{22}$~cm$^{-2}$ in the cloud, we also find that all five sub-regions are magnetically subcritical. These results indicate that the magnetic field in L183 is sufficiently strong to oppose the gravitational collapse of the cloud.

\end{abstract}


\section{Introduction} \label{sec:intro}

In the current paradigm of star formation, the filamentary structures found in molecular clouds are expected to fragment into dense cores of dust and gas (with hydrogen volume densities of $n_{\text{H}_2}>10^4$~cm$^{-3}$) as a necessary step before stars can be formed through gravitational collapse. Indeed, far-infrared and submillimeter observations in the past two decades have shown that these dense cores are ubiquitous in nearby star-forming regions \citep[e.g.,][]{Andre2014}. However, not all cores are observed to harbor a protostar \citep[e.g.,][]{difrancesco2007}. These \enquote{starless} cores are typically divided into two categories: (1)unbound cores supported against gravity by thermal pressure, which, along with gravitationally bound cores \citep[such as B86][]{2001Natur.409..159A}, can be modeled as Bonnor-Ebert spheres \citep{Ebert1955,1956MNRAS.116..351B}, and (2) collapsing prestellar cores transitioning into first hydrostatic cores\footnote{These objects represent an early phase in the low-mass star formation process, after collapse of the parent core has begun but before a true protostar has formed.} \citep[e.g.,][]{Machida2008}.

\begin{figure}
	\includegraphics[width=\columnwidth]{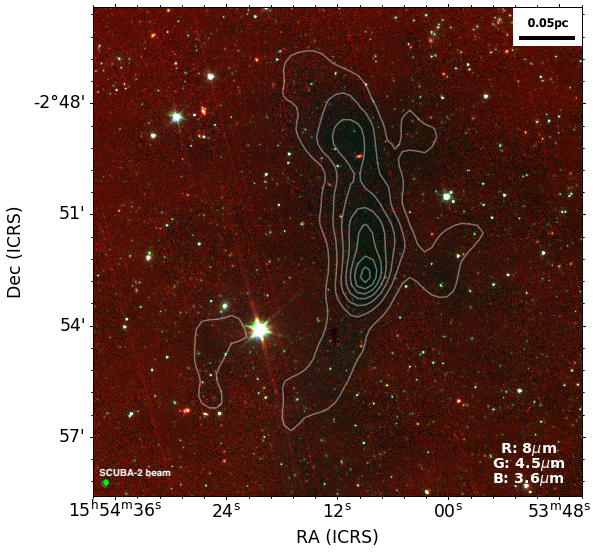}
	\caption{Combined \textit{Spitzer} observations of L183 taken with IRAC at 8~$\micron$, 4.5~$\micron$, and 3.6~$\micron$. The contours trace the 850~$\micron$ dust emission map from SCUBA-2 starting at 10 mJy~beam$^{-1}$ and increasing in increments of 20 mJy~beam$^{-1}$. The green circle in the lower left corner shows the JCMT/SCUBA-2 beam size.}
	\label{Fig:struct}
\end{figure}

The L183 cloud \citep[][aka L134N]{lynds1962}, and its starless cores ($Spitzer$ image shown in Figure \ref{Fig:struct}), is an ideal candidate to study the role of magnetic fields at the onset of star formation, and specifically to probe if they can moderate the gravitational collapse of pre-stellar cores. Indeed, at a distance of 110 pc \citep{1989A&A...223..313F}, the proximity and low Galactic longitude ($l=6.1$) of L183 means it is a cloud with a significant number of background stars despite its high Galactic latitude ($b=36.8$) \citep{Pagani2003}.

\citet{Lee1999,Lee2001,Lee2004} listed L183 as a possible infall candidate based on spectroscopic measurements of the CS~(2$-$1), CS~(3$-$2), N$_2$H$^+$~(1$-$0), and DCO$^+$~(2$-$1) molecular lines. The L183 core has a C$^{18}$O depletion level typically associated with chemically-evolved cores \citep{Tafalla2005a, Tafalla2005b}, yet it shows no signs of hosting embedded young stellar objects even though less-evolved cores like L1521F are already undergoing star formation (\citealt{Tafalla2005a, Tafalla2005b}; Soam et al. in prep.). It is therefore possible that the gravitational collapse of L183 is significantly curtailed either because of the gas kinematics or the magnetic energy inside the core.

In this work, we investigate the contribution of the magnetic field to the stability of L183. This is achieved by using 850~$\mu$m observations obtained with the POL-2 polarimeter at the James Clerk Maxwell Telescope (JCMT). The structure of magnetic fields in the interstellar medium can be directly inferred from the polarization of dust thermal emission at far-infrared and submillimeter wavelengths \citep[see][and references therein]{ALV2015}. Such emission polarization is expected to be perpendicular to the plane-of-the-sky field orientation due to the alignment of interstellar dust grains with magnetic fields through Radiative Alignment Torques (RATs). The alignment efficiency of dust grains in the dense environment of L183 will be investigated in a forthcoming paper (Andersson \& Soam et al. in prep.).

\begin{figure}
    \includegraphics[width=\columnwidth]{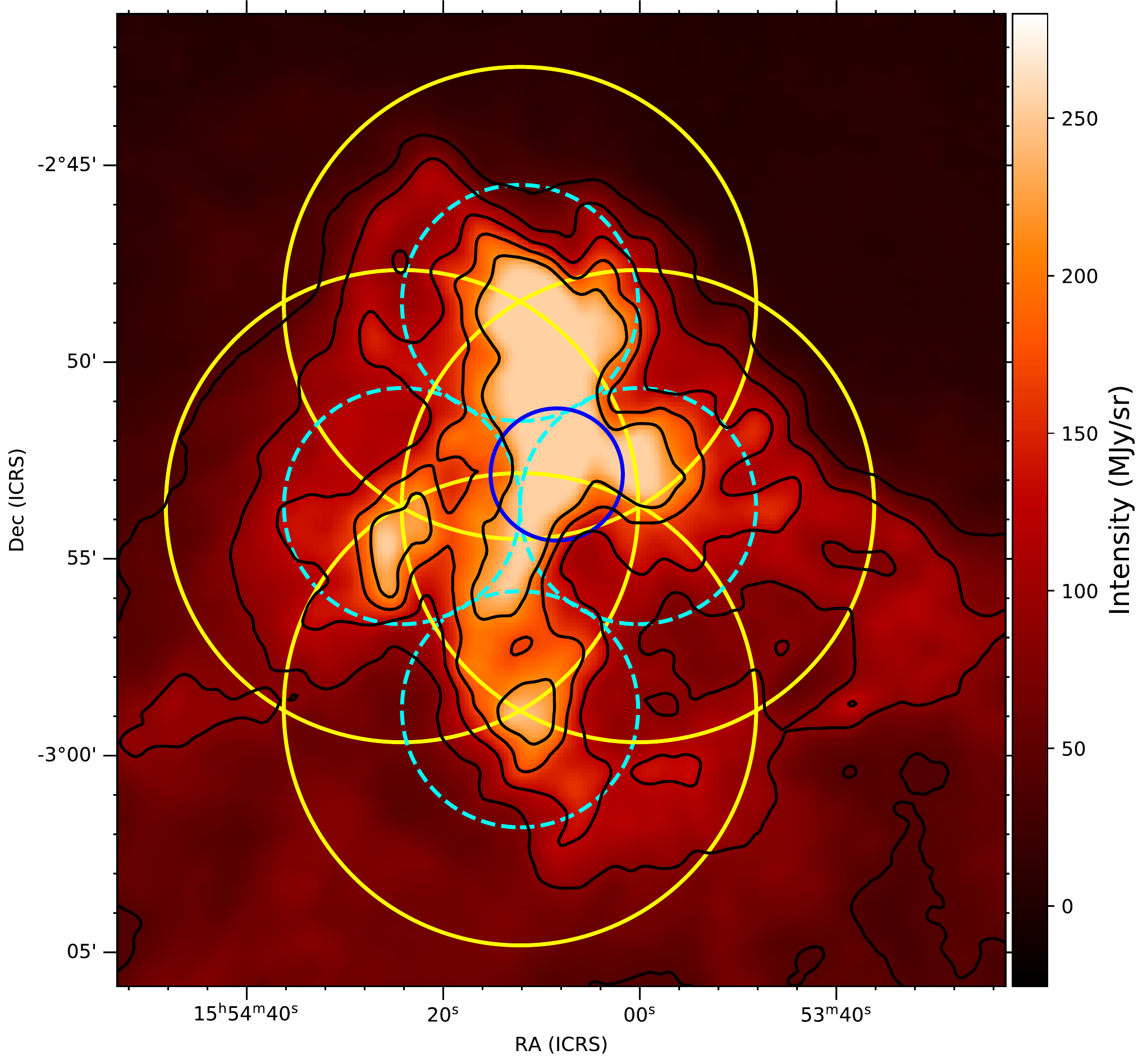}
	\caption{The dust emission toward L183 as traced by \textit{Herschel} 500~$\mu$m SPIRE observations. The plain blue circle in the center indicates the limited area previously observed with SCUPOL by \citet{2004Ap&SS.292..225C}. The yellow circles indicate the four POL-2 Daisy fields covered in our observations. These field$-$of$-$views are optimized so that their central $6\arcmin-$wide areas (dashed cyan circles), where POL-2's sensitivity is optimal, are centered on high density regions while still fully covering the main body of the L183 cloud.}
	\label{Fig:proposed}
\end{figure}

Assuming a distance of 110~pc, our 850~$\mu$m observations achieve a spatial resolution of $7.5$~mpc (or 1600~au) while simultaneously mapping all of the highest extinction regions in the cores over a $\sim 12$\arcmin-wide field (see Figure~\ref{Fig:struct} and \ref{Fig:proposed}). \citet{2004ApJ...600..279C} also observed L183 at a comparable resolution using the SCUPOL polarimeter (the predecessor of POL-2) at the JCMT, but they were limited to a much smaller $\sim 3.5$\arcmin-wide region with a lower sensitivity than our POL-2 data (see Section~\ref{sec:comp}). Nevertheless, their original analysis suggested that L183 may be only weakly supercritical, i.e., the magnetic energy is at least three times smaller than, and at most equal to, the gravitational energy in the core ($E_{grav} /3 < E_{mag} < E_{grav}$). With the higher sensitivity of POL-2, we significantly improve the accuracy of this criticality measurement. 

The POL-2 data presented in this work provide the deepest polarization observations to date of a starless core. Although significant improvements have been made in recent years to the sensitivity of polarimetric instruments, polarimetry at far-infrared and submillimeter wavelengths still presents unique technical challenges \citep[see][]{PattleFissel2019} that are compounded by the faint polarization signature of starless cores. This low polarized emission is explained by the combination of two main factors: First, the dust content of starless cores is typically colder than in active star-forming regions \citep[$T_d \approx 7$~K for L183,][see also Section~\ref{sec:colden}]{Pagani2003}, thus leading to weaker dust thermal emission. Second, the polarization efficiency in starless cores is known to decrease sharply as a function of visual extinction $A_V$ \citep[e.g.,][]{Alves2014,Jones2015}, which results in a decrease of the degree of polarization in the denser parts of the cloud. Thus, observations to date have been used mostly to study magnetic fields in all but the brightest starless cores. Prior to this work, \citet{2004ApJ...600..279C} and \citet{2000ApJ...537L.135W} have used SCUPOL to study the bright starless cores L183, and L1544 and L43, respectively.

This paper is structured as follows: Section~\ref{sec:obs} presents the observations and the data reduction process. Section~\ref{sec:analysis} provides a discussion of the main results, such as the dust properties (\S\ref{sec:colden}), the magnetic field morphology (\S\ref{sec:morpho}), the field strength (\S\ref{sec:stren}), the criticality criterion (\S\ref{sec:critic}), and the energy budget of the cloud (\S\ref{sec:ener}), as well as a comparison with previous SCUPOL results (\S\ref{sec:comp}). Finally, we summarize the findings of this paper in Section~\ref{sec:summary}.

\section{Observations and data reduction} \label{sec:obs}

The observations were conducted with SCUBA-2/POL-2 at 850\,$\mu$m in February and March of 2019 (M19AP009; PI: Bastien, P.) using the polarimetric daisy-map mode of the JCMT (\citealt{2013MNRAS.430.2513H, 2016SPIE.9914E..03F}; P.~Bastien et al. in prep.). The POL-2 polarimeter, which consists of a fixed polarizer and a half-wave plate rotating at a frequency of 2~Hz, is placed in the optical path of the SCUBA-2 camera. Figure \ref{Fig:proposed} shows the locations of the observations on L183 \textit{Herschel}/SPIRE image. The weather conditions during observations were split between $\tau_{225} < 0.05$ and $0.05< \tau_{225} < 0.08 $, where $\tau_{225}$ is the atmospheric opacity at 225~GHz. The total integration time for a single field was $\sim$4 hours to complete six full daisy patterns. SCUBA-2/POL-2 simultaneously collects data at 450~$\mu$m and 850~$\mu$m with effective full-width half-maximum (FWHM) beam sizes of 9$\farcs$6 and 14$\farcs$1, respectively \citep{2013MNRAS.430.2534D}. For this work, we focused exclusively on the 850~$\mu$m data due to the signal-to-noise ratio (SNR) in the 450~$\mu$m data being too low to recover a sufficient number of polarization vectors for the analysis. 

We observed four separate sub-regions (North, South, East, and West) overlapping near the center of the cloud. For the POL-2 daisy-map mode, a fully sampled circular region of 12$\arcmin$ diameter is produced, with a high signal-to-noise coverage over the central 6$\arcmin$ wide area. This observing mode is based on the SCUBA-2 constant velocity daisy scan pattern \citep{2013MNRAS.430.2513H}, but modified to have a slower scan speed (i.e., 8$\arcsec$~s$^{-1}$ compared to the original 155$\arcsec$~s$^{-1}$) to obtain sufficient on-sky data to measure the Stokes \textit{Q} and \textit{U} values accurately at every point of the map. The integration time decreases toward the edges of the map, which consequently leads to an increase in the root mean square (RMS) noise levels. 

To reduce the data, we used the STARLINK/SMURF \citep{2013MNRAS.430.2545C, 2014ASPC..485..391C} package {\tt\string pol2map} specifically developed for submillimeter data obtained with the JCMT. The details of the data reduction procedure are presented in \citet{2019ApJ...876...42W}, and we will only summarize the relevant steps here. 

First, the raw bolometer time-streams are converted into Stokes \textit{I}, \textit{Q}, and \textit{U} time-streams at a sampling rate of a full half-wave plate rotation through the process {\tt\string calcqu}. A Stokes~\textit{I} map is then created from all Stokes \textit{I} time-streams using the routine {\tt\string makemap}, which is an iterative map-making process. Individual \textit{I}~maps corresponding to each observation were co-added to produce the initial \textit{I} map of each region \citep[cf][]{2013MNRAS.430.2545C}. Because four separate regions were observed to cover the cloud, we co-added the initial Stokes~\textit{I} map from each region to get the complete Stokes~\textit{I} map. 

The final Stokes~\textit{I}, \textit{Q}, and \textit{U} maps were obtained by running {\tt\string pol2map} a second time. The initial Stokes~\textit{I} mosaic map is used to generate a fixed SNR-based mask for all subsequent iterations of {\tt\string makemap}. During the final process, we corrected for the loss of synchronization between data values and pointing information in the data reduction process via the {\tt\string skyloop}\footnote{http://starlink.eao.hawaii.edu/docs/sc22.pdf} parameter in {\tt\string pol2map}. This parameter improves the recovery of fainter, extended emission in the map by iterating individual observations in parallel. This is in contrast to the traditional JCMT map-making method of deriving an iterative solution for each observation individually.   The resulting Stokes~\textit{I}, \textit{Q}, and \textit{U} maps for the four separate regions were then co-added to produce the final maps used for this study. 

The resulting Stokes~\textit{I}, \textit{Q}, and \textit{U} maps were flux calibrated, in units of $\rm mJy\,beam^{-1}$, using a Flux Calibration Factor (FCF) for 850\,$\mu$m of 725 Jy\,$\rm pW^{-1}$\footnote{This conversion was done using the CALIBRATE-SCUBA2-DATA recipe under the PICARD package in STARLINK}. The final co-added Stokes \textit{I}, \textit{Q}, and \textit{U} maps have an RMS noise\footnote{This value was measured using the SCUBA2-MAPSTATS recipe under the PICARD package in STARLINK} of $\rm\sim\,1.5\,mJy\,beam^{-1}$. Finally, the data were reduced with a 12$\arcsec$ pixel size at each step.

After the final step of running {\tt\string pol2map}, we obtain a polarization vector catalog produced from the co-added Stokes \textit{I}, \textit{Q}, and \textit{U} maps. The final polarization values obtained here are debiased using the Stokes \textit{Q} and \textit{U} variances to remove the statistical bias in regions of low signal-to-noise ratio (SNR) \citep{1974ApJ...194..249W}. 

The values for the debiased degree of polarization $P$  were calculated from 

\begin{equation}
P=\frac{1}{I}\sqrt{Q^{2}+U^{2}-\frac{1}{2}(\delta Q^{2}+\delta U^{2})}   \,\, ,
\end{equation}
where \textit{I}, \textit{Q}, and \textit{U} are the Stokes parameters, and $\delta Q$, and $\delta U$ are the uncertainties for Stokes \textit{Q} and \textit{U} (see \citet{2019ApJ...883...95S, 2019ApJ...876...42W}). The uncertainty $\delta P$ of the polarization degree was obtained using

\begin{equation}
\delta P = \sqrt{\frac{(Q^2\delta Q^2 + U^2\delta U^2)}{I^2(Q^2+U^2)} + \frac{\delta I^2(Q^2+U^2)}{I^4}}  \,\, ,
\end{equation}
with $\delta I$ being the uncertainty for the Stokes~\textit{I} total intensity. 

The polarization position angles $\theta$, increasing from north to east in the sky projection, were measured using the relation 

\begin{equation}
{\theta = \frac{1}{2}tan^{-1}\frac{U}{Q}} \, .
\end{equation}

The corresponding uncertainties in $\theta$ were calculated using

\begin{equation}
\delta\theta = \frac{1}{2}\frac{\sqrt{Q^2\delta U^2+ U^2\delta Q^2}}{(Q^2+U^2)} \times\frac{180\degree}{\pi}  \,\, .
\label{eq:dtheta}
\end{equation}

The plane-of-sky orientation of the magnetic field is inferred by rotating the polarization angles by 90$\degree$ (assuming that the polarization is caused by elongated dust grains aligned perpendicular to the magnetic field). 


\section{Analysis} \label{sec:analysis}
\subsection{Polarization and Magnetic fields}

\begin{figure*}
	\centering
	\resizebox{16.8cm}{10cm}{\includegraphics{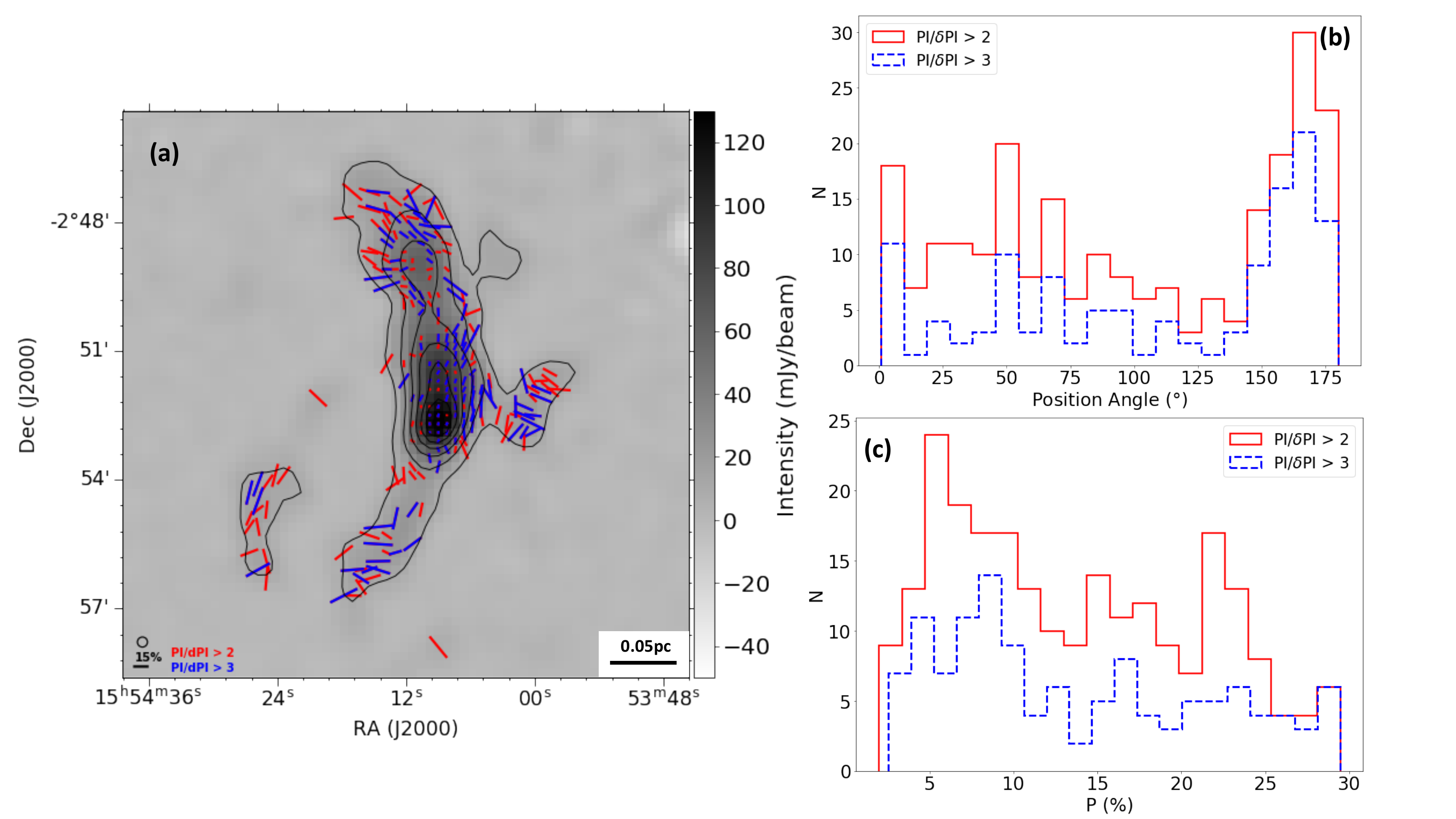}}
	\caption{Panel (a) shows the magnetic field orientations in L183 with $PI/\delta PI > 2$ (red vectors) and $PI/\delta PI> 3$ (blue vectors). The vectors are plotted on the 850\,$\mu$m dust emission map with overlaid contours starting at 10 mJy~beam$^{-1}$ and increasing by 20 mJy~beam$^{-1}$. The JCMT 850\,$\mu$m beam size and the vector scale are shown in the bottom left corner. Panels (b) and (c) show the distributions of magnetic field position angles and polarization percentages, respectively.}
	\label{Fig:SN}
\end{figure*}

In our analysis of the POL-2 data, we only use data points where the observed uncertainties in position angle are less than 20$\degree$. Additionally, we impose an additional constraint of $I/\delta I > 10$ to improve the reliability of our analysis. We checked the quality of the data used for the analysis by examining different SNR values derived from the polarization intensity (\textit{PI}) and its uncertainty ($\delta_{PI}$). In Figure \ref{Fig:SN}a, the magnetic field orientations inferred from SNR $> 3$ ($PI/\delta_{PI} > 3$; 124 blue vectors) and SNR $>$ 2 ($PI/\delta_{PI} > 2$; 236 red vectors [some of the red vectors are hidden under the blue ones]) are generally consistent within the cloud. The goal of this comparison is to evaluate the validity of using the lower SNR threshold of SNR $>$ 2 instead of the stricter SNR $>$ 3. The additional vectors plotted using SNR $> 2$ have similar polarization percentages and position angles to the SNR $> 3$ vectors in their vicinity, which suggests that the larger population of SNR $>$ 2 vectors can be used for the analysis. Additionally, in panels (b) and (c) of Figure \ref{Fig:SN}, the distributions of position angles and polarization percentages follow similar behaviors, further justifying our use of the $PI/\delta_{PI} > 2$ values.

\begin{figure}
	\centering
	\resizebox{8.5cm}{11.0cm}{\includegraphics{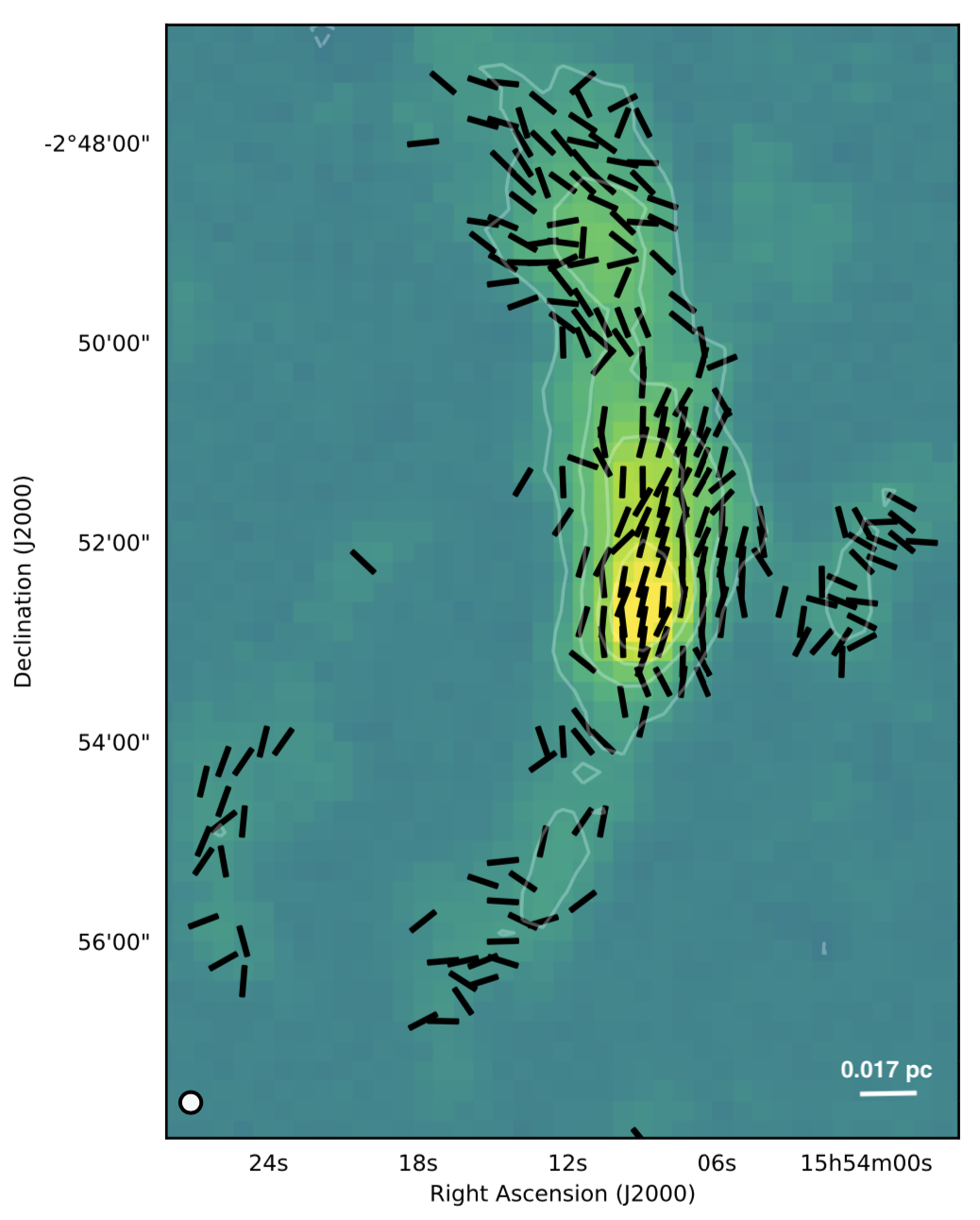}}
	\caption{Magnetic field orientations in L183 obtained after rotating the polarization vectors by 90$\degree$ and shown as normalized line-segments independent of the polarization degree $P$. These vectors correspond to data with $PI/\delta_{PI} > 2$, where \textit{PI} and $\delta_{PI}$ are respectively the polarized intensity and its uncertainty. The JCMT beam size at 850~$\mu$m is shown in the lower left corner.}\label{Fig:Bnorm}
\end{figure}

\begin{figure*}
	\centering
	\resizebox{18cm}{10.12cm}{\includegraphics{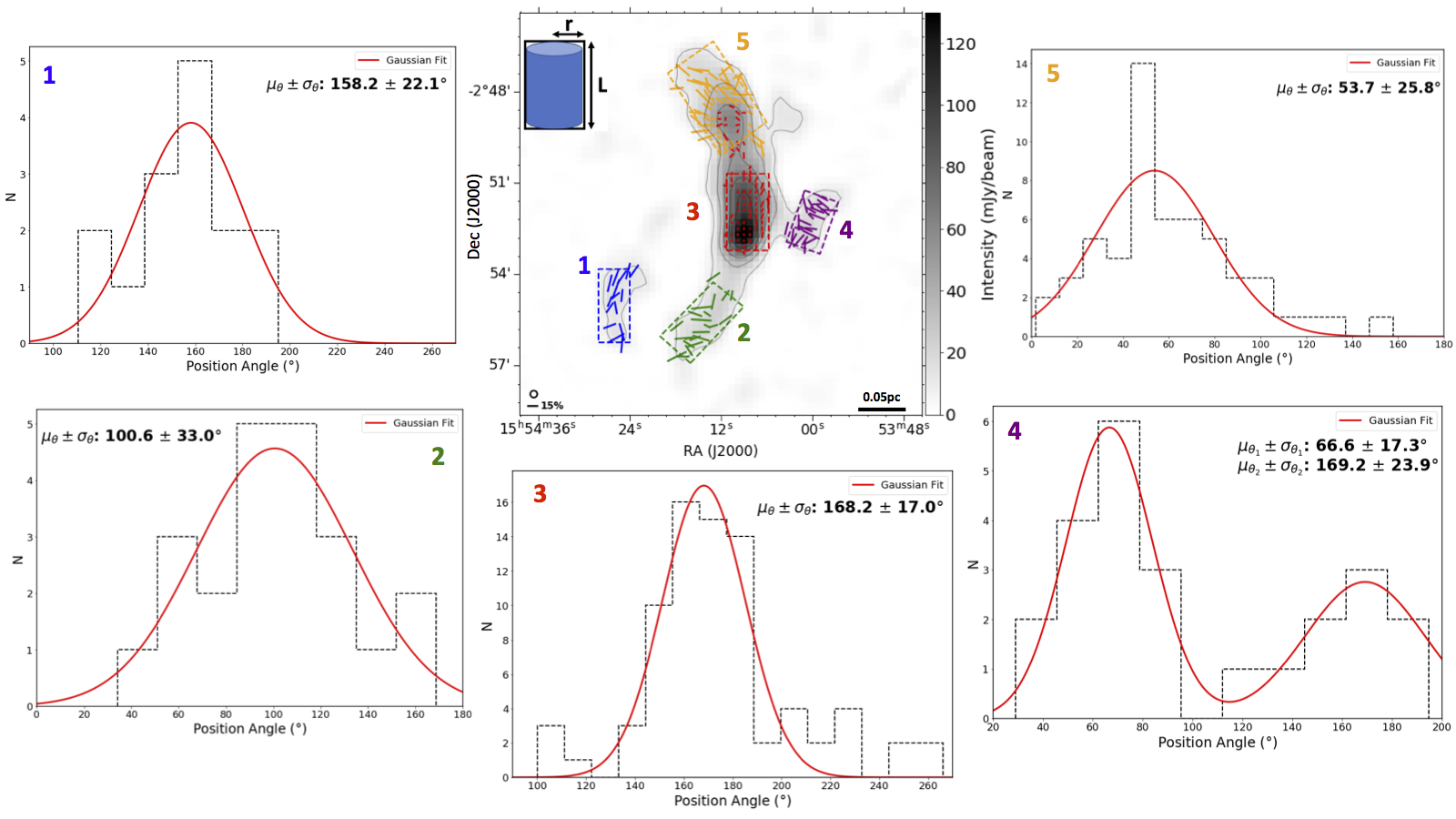}}
	\caption{The five regions identified in L183 with their respective distributions of position angles. Each panel includes the mean and standard deviation of the Gaussian fit (red line). The background image is the 850$\micron$ dust emission map. The scale for the vectors is shown in the bottom left along with JCMT 850$\micron$ beam size. The red vectors show data with $I > 50$mJy~beam$^{-1}$, while all other vectors instead show data where $I < 50$mJy~beam$^{-1}$. The regions are drawn as rectangular boxes with dimensions listed in Table \ref{tab:main} as shown by the the dust emission map.}
	\label{Fig:reg}
\end{figure*}

Figure \ref{Fig:Bnorm} shows the morphology of the magnetic field in the inner parts of the L183 cloud. Here, the lengths of the vectors have been normalized for clarity; they do not represent the polarization percentage. Rather than evaluating the magnetic field structure of L183 by using a single structure function, we instead chose to split the cloud into five regions with distinct populations of magnetic field lines based on their apparent uniformity in polarization angle (see Figure \ref{Fig:reg}). This allows us to employ the classical interpretation of the Davis-Chandrasekhar-Fermi method (DCF; \citealt{1953ApJ...118..113C}) to derive the magnetic field strength independently for each region. Previous studies of magnetic field strength \citep[e.g.][]{2019ApJ...877...88C}, use an improved DCF method developed by \citet{Houde_2009} and \citet{Hildebrand_2009} which uses an angular dispersion function. We used this angular dispersion function on the five regions seen in Figure \ref{Fig:reg}. However, for regions 1 through 4, the function failed to converge in 200 iterations. In region 5, the calculated turbulent-to-ordered magnetic energy ratio, $\langle B_{t}^{2}\rangle/\langle B_{o}^{2}\rangle$, was 21.93$\pm$46.56 and the reduced chi-squared value for the fit of the angular dispersion function was 5.3. The large energy ratio uncertainty gives an indeterminate result. For these observations, the data sets are too noisy and therefore the contribution from turbulence cannot be distinguished from beam smoothing. As a reference, our noise level ($\rm\sim\,1.5\,mJy\,beam^{-1}$) is comparable to that of SCUBA-2/POL-2 observations of B1 \citep{2019ApJ...877...88C} although L183 is approximately 3 times dimmer. We therefore use the DCF method used in \citet{2004ApJ...600..279C} to calculate the magnetic field strength.

\begin{figure*}
	\centering
	\resizebox{12cm}{8.4cm}{\includegraphics{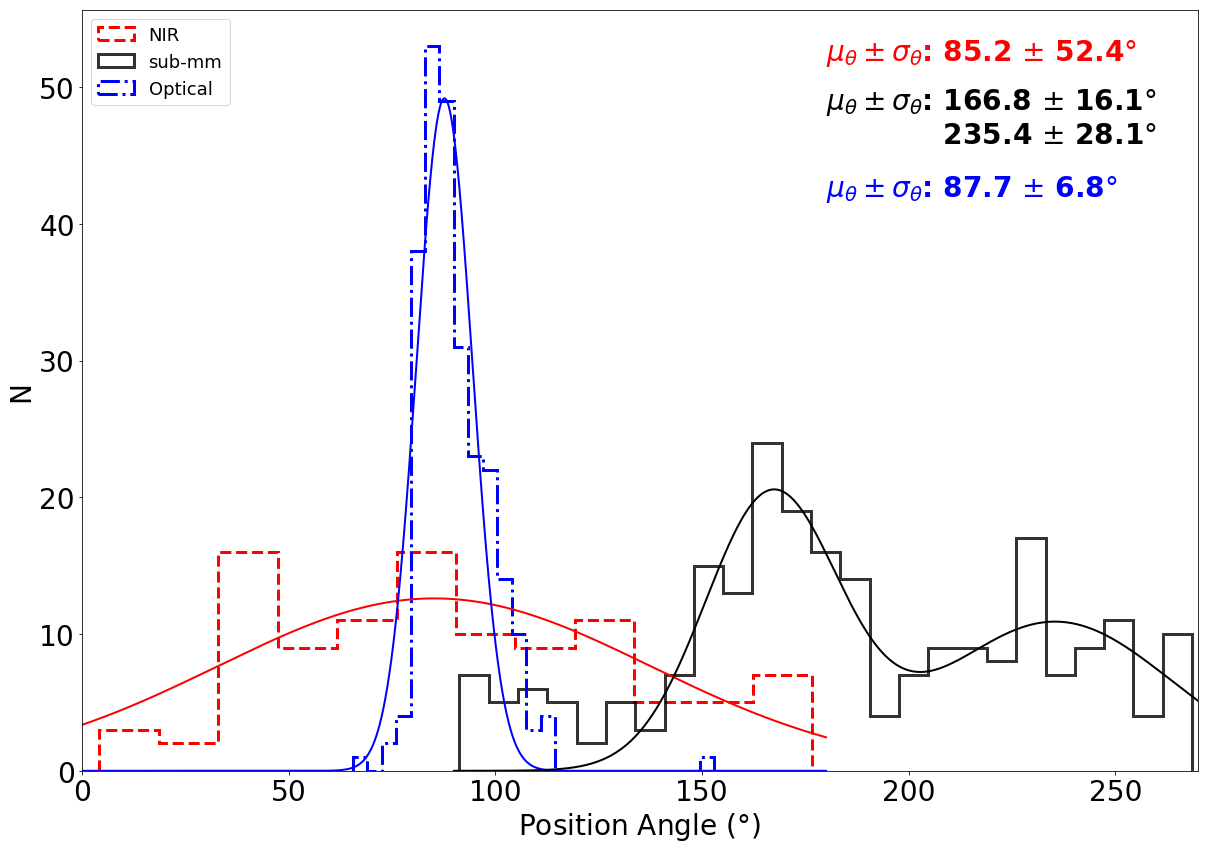}}
	\caption{Histogram showing the distributions of magnetic field position angles (POL-2 polarization position angles rotated by 90$\degree$). Due to the 180$\degree$ ambiguity in magnetic field direction, the POL-2 position angles that are less than 90$\degree$ have been rotated by 180$\degree$ to best demonstrate the observed double-Gaussian distribution. The POL-2 position angles correspond to data with $PI/\delta PI > 2$.}
	\label{Fig:dist}
\end{figure*}

Figure \ref{Fig:reg} shows the five regions and their respective distributions of position angles, including the Gaussian fit for each. The total distribution of position angles shown in Figure \ref{Fig:dist} appears to follow a double Gaussian distribution, although the second peak is rather broad. In regions 1, 3, and 4, we have taken advantage of the 180$\degree$ ambiguity in magnetic field direction (e.g., a vector with position angle of 15$\degree$ shows the same direction as one with a position angle of 195$\degree$) to best demonstrate the Gaussian distributions of the position angles. 

The DCF method assumes that the geometry of the magnetic field is uniform in each region, and so measuring the dispersion of position angles allows us to estimate the field strength. This dispersion in position angles is explained by local turbulence disrupting the magnetic field structure. We also assume that the distribution of vectors around the mean field direction is approximately Gaussian, and is therefore well-characterized by its standard deviation. The DCF method determines the field strength using following equation:

\begin{equation}
B_{pos} = Q_{c}\sqrt{4\pi\rho}\frac{\sigma_{v}}{\sigma_{\theta}} \;,
\label{eq:dcf}
\end{equation}
where $Q_{c}$ is a correction factor that accounts for variations of the magnetic field on scales smaller than the beam, $\rho$ is the gas density, $\sigma_{v}$ is the one-dimensional non-thermal velocity dispersion of the gas, and $\sigma_{\theta}$ is the dispersion in polarization angle. \citet{2004Ap&SS.292..225C} further approximated this formulation as

\begin{equation}
B_{pos} \approx 9.3\sqrt{n(H_{2})}\frac{\Delta v}{\sigma_{\theta}} \mu G \;,
\label{eq:dcf_sim}
\end{equation}
where $Q_{c}$ has been taken to be 0.5 \citep{2001ApJ...546..980O}, $n(H_{2})$ is volume density of molecular hydrogen, and $\Delta v$ is the FWHM of the gas velocity calculated by $\Delta v$ = $\sigma_{v}\sqrt{8\rm{ln}2}$. The units are $\rm cm^{-3}$ for the volume density $n(H_{2})$, $\rm km~s^{-1}$ for $\Delta v$, and degrees for $\sigma_{\theta}$.

\subsection{Dust column densities and temperatures}\label{sec:colden}

\begin{figure*}
	\centering
	\resizebox{18.5cm}{7.5cm}{\includegraphics{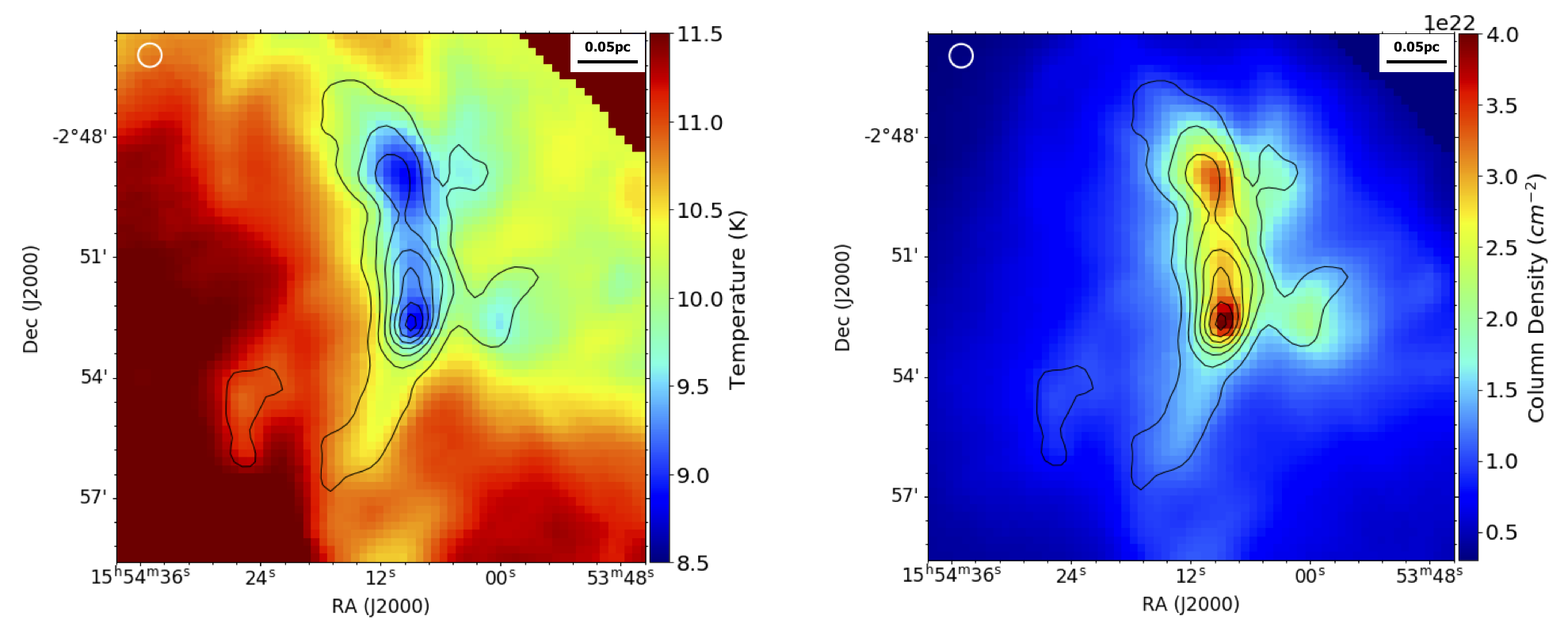}}
	\caption{The left panel shows the fitted dust temperature $T_d$ map in L183, and the right panel shows the corresponding column density $N_{\text{H}_2}$ map. The overlaid contours on both plots are from the SCUBA-2 850$\mu$m dust emission map. Each map is smoothed to the 35$\farcs$2 beam size of the \textit{Herschel} 500~$\mu$m observations, which is shown by the circle in the upper left corner.}
	\label{Fig:cd}
\end{figure*}

L183 was previously observed by the \textit{Herschel} space observatory with the Spectral and Photometric Imaging Receiver (SPIRE) at 250, 350, and 500 $\micron$, as well as with the Photodetector Array Camera \& Spectograph (PACS) at 100 and 160 $\micron$. We use this archival \textit{Herschel} data at 160, 250, 350, and 500 \micron, combined with our JCMT data at 850~\micron, to fit a modified black-body function (see Equation~\ref{eq:mod}) for the dust emission in L183. The \textit{Herschel/PACS}, \textit{Herschel/SPIRE}, and JCMT 850 $\micron$ images were smoothed to the SPIRE 500 $\micron$ FWHM beam size of 35$\farcs$2 and then re-projected on a common grid. 

The spectral energy distribution (SED) for each pixel was fitted assuming the following formula for a modified black-body emission \citep[see][]{2008A&A...487..993K}:

\begin{equation}
S_\nu = B_\nu (T_\mathrm{d})(1-\mathrm{e}^{-\tau_\nu}) \;,
\label{eq:mod}
\end{equation}
\begin{equation}
B_\nu(T_\mathrm{d}) = \frac{2h\nu^3}{c^2}\frac{1}{\mathrm{e}^{h\nu/k_\mathrm{B}T_\mathrm{d}}-1} \;,
\label{eq:BB}
\end{equation}
\begin{equation}
\tau_\nu = \mu_\mathrm{H_2} m_\mathrm{H} \kappa_\nu N_\mathrm{H_2} \;,
\label{eq:tau}
\end{equation}
and
\begin{equation}
\kappa_\nu = \kappa_o \, \left( \frac{\nu}{\nu_o} \right) ^\beta \;, 
\label{eq:kap}
\end{equation}
where $S_\nu$ is the measured flux at the observed frequency $\nu$, $B_\nu(T_\mathrm{d})$ is the Planck function for a dust temperature $T_\mathrm{d}$, $\tau_\nu$ is the optical depth, $\mu_{H_2}$ is the mean molecular weight of the hydrogen gas in the cloud, $m_{H}$ is the mass of an hydrogen atom, $N_{\text{H}_2}$ is the column density, and $\kappa_\nu$ is the dust opacity (absorption coefficient). We use a value of 2.8 for $\mu_{H_2}$, and $\kappa_\nu$ was calculated for each frequency observed using Equation \ref{eq:kap}, where $\beta$ is the emissivity spectral index of the dust, and we assume $\kappa_o = 0.1$~cm$^2$~g$^{-1}$ and $\nu_{o} = 10^{12}$\ Hz  \citep{1990AJ.....99..924B}. 

The SED fitting for each pixel was completed in two steps. In the first step, $\beta$ was left as a free parameter to be fitted simultaneously with the temperature $T_d$ and the column density $N_{\text{H}_2}$. In the second step, we instead fixed $\beta$ to the best-fit value obtained from the first step, before re-doing the SED fit to obtain the final values for the temperature and column density. The temperature and column density maps obtained through this procedure for L183 are shown in Figure~\ref{Fig:cd}.

The dust temperatures (left panel of Figure \ref{Fig:cd}) in the filament vary approximately between 8.8 and 11~K, with very cold dust present in the two central cores of L183. These results are consistent with those of \citet{2002MNRAS.329..257W}, who found a temperature of $10 \pm 3$~K in the main, southern, core by fitting a modified black-body curve to ISOPHOT measurements, as well as with those from \citet{2003A&A...398..571L}, who found a colour temperature of $8.3 \pm 0.4$~K. 

The derived column densities (right panel of Figure \ref{Fig:cd}) peak at $\sim 4 \times 10^{22}$ $\rm cm^{-2}$ in the main core. The average column density in this core is around $3.0 \times 10^{22}$ cm$^{-2}$, which agrees with the average value of $2.7 \times 10^{22}$ cm$^{-2}$ found by \citet{2004ApJ...600..279C}. 

We estimated the hydrogen volume densities $n(\text{H}_2)$ of the five regions identified in Figure~\ref{Fig:reg} by assuming they each have a cylindrical geometry, and by adopting the same procedure as \citet{2018ApJ...859..151L}. The projected lengths $L$ and radii $r$ of the cylinders for each region of L183, as well as their estimated volume densities and total masses, are given in Table \ref{tab:main}.

\subsection{Magnetic field morphology}\label{sec:morpho}

\begin{figure*}
	\centering
	\resizebox{17.5cm}{9.5cm}{\includegraphics{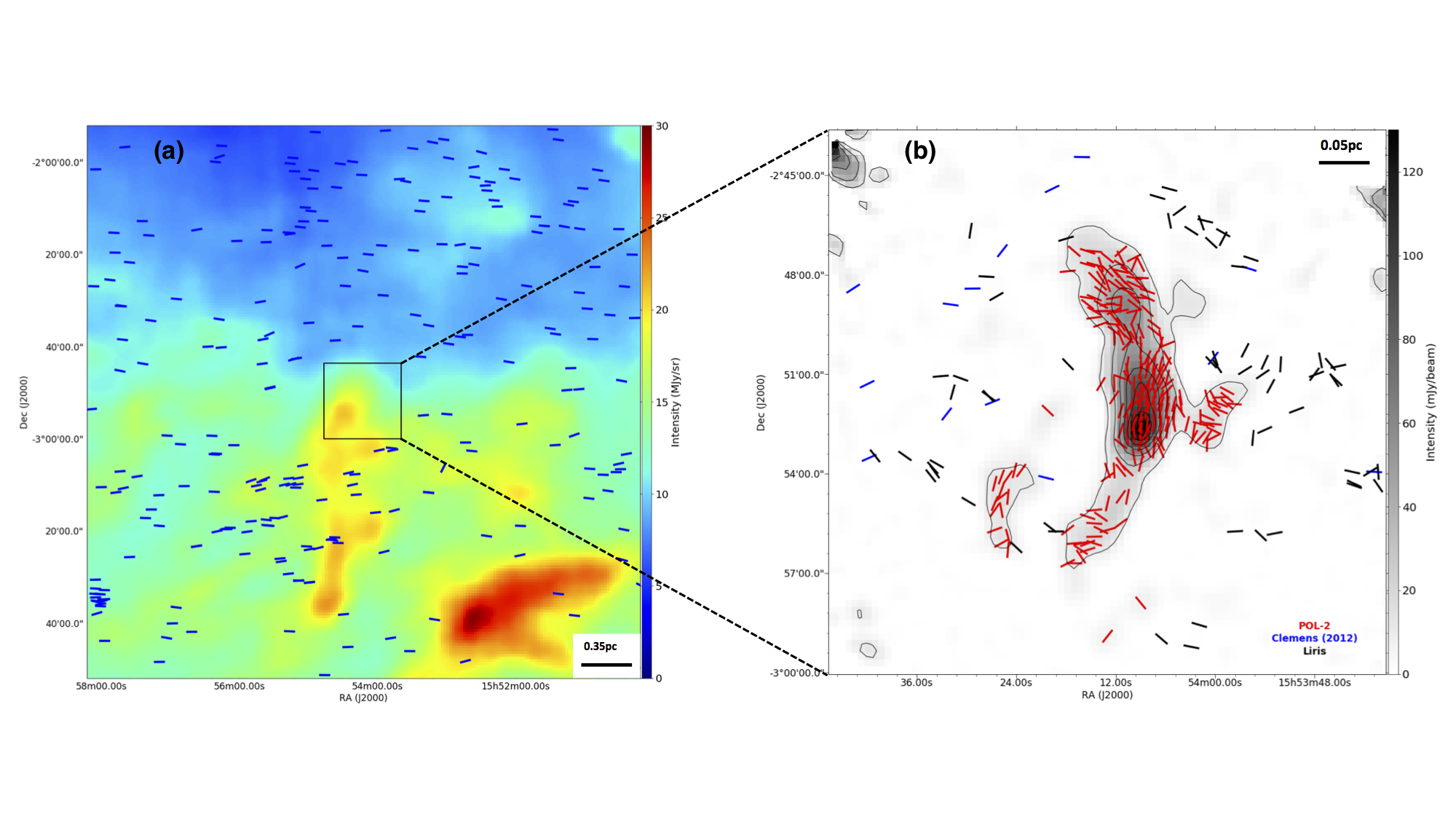}}
	\caption{Panel (a): R-band polarization vectors acquired with the 2.1\,m telescope at McDonald observatory (Andersson \& Soam et al., in prep.) plotted on the 100~$\micron$ IRAS emission map. The vector lengths are normalized and do not reflect polarization percentages. The distance scale for the image is shown in the lower right corner. The inset is the location of the L183 cloud observed at 850~$\micron$ by SCUBA-2/POL-2 at the JCMT, and the zoomed-in image is shown in panel (b). Panel (b): H-band polarization measurements from LIRIS at the William Herschel Telescope (Andersson \& Soam et al., in prep.) are shown in black, while H-band polarimetric data from the Mimir instrument at the Perkins telescope \citep{2012ApJ...748...18C} are shown in blue. The 850~$\micron$ polarization observations (this work) are shown with normalized red vectors that have been rotated to show the orientation of the magnetic field, consistent with the optical and NIR vectors. The contours follow the 850~$\micron$ dust emission. The emissions at 100~$\micron$ and 850~$\micron$ microns do not peak at the same location.}
	\label{Fig:scale}
\end{figure*}

Significant amounts of complimentary optical and near-infrared (NIR) polarization data exist for L183 \citep[][Andersson et al., in prep.]{2012ApJ...748...18C}. Panel (a) of Figure \ref{Fig:scale} shows the optical polarization vectors overlaid on a 100~$\micron$ IRAS map. Panel (b) shows the zoomed-in area of panel (a) with the NIR and submillimeter polarization vectors overlaid on the JCMT 850~$\micron$ dust emission map. Figure \ref{Fig:dist} shows the distribution of polarization position angles for all of the data sets, optical, NIR, and submillimeter. Because of the 180$\degree$ ambiguity in polarization position angles we have, for the submillimeter polarization, added 180$\degree$ to any angle less than 90$\degree$ in order to make the position angle distribution continuous. Hence, the range in submillimeter wave polarization position angles is 90$\degree$ to 270$\degree$.

The optical and NIR polarization data both show single peak distributions averaging around 90$\degree$, meaning that the large-scale magnetic field is oriented in the east-west direction (Figure \ref{Fig:scale}). However, the polarization within the core is very different from the large-scale orientation, with a double-peaked distribution instead. We find a distinct peak for the submillimeter position angles around 180$\degree$ with an additional broad distribution centered around 230$\degree$. As seen in Figure~\ref{Fig:scale}, the former corresponds with the magnetic field in the main core, and is perpendicular to the large-scale field. The broader distribution has a mean of 235$\degree$ (or 55$\degree$) which still differs from the large-scale orientation. \citet{sharma2018} also reported the  east-west orientation of B-fields in L183 using V-band polarization measurements.

Planck all-sky polarization measurements found interstellar magnetic fields mostly parallel to the diffused low-density regions of the filamentary molecular clouds, whereas field lines tend to be perpendicular in high-density regions \citep{2015A&A...576A.104P, 2016A&A...586A.138P}. Several studies on elongated infrared dark clouds (IRDCs) have also seen similar features. For instance, \citet{2019ApJ...883...95S} and \citet{2018ApJ...859..151L} have seen changing magnetic fields orientations from the diffuse to dense regions of the G34.43+0.24 and G035.39-00.33 clouds, respectively. It is hard to see a clear change in the magnetic field orientation within the filament containing L183, but it can be noticed that magnetic field lines are following the shape of the filament in diffuse parts such as in regions 1 and 5 (see Figure \ref{Fig:reg}), but not completely parallel to the long axis of filamentary part in region 2. Whereas, in the dense core (region 3), the magnetic field lines are not perpendicular to the filament either but are instead parallel to it.

\subsection{Magnetic field strength}\label{sec:stren}

The magnetic field strength in each of the regions identified in L183 (see Figure~\ref{Fig:reg}) can be calculated with Equation~\ref{eq:dcf_sim}. As described in Section~\ref{sec:colden}, the volume density for each region can be found in Table~\ref{tab:main}. Additionally, the values for $\Delta v$ were taken from $\rm N_{2}H^{+}$(1-0) measurements by \citet{Lee2001}. Furthermore, we employed a similar method to the one presented by \citet{2004ApJ...600..279C} in order to calculate the dispersion of polarization angles $\sigma_{\theta}$. 

Within each region, we measure the difference between the polarization angle at a given position and the mean polarization angle in the region (i.e., $\theta - \bar{\theta}$), which should probe the random variations of the magnetic field. The distribution of these measured deviations was fitted with a Gaussian function, and we combined the mean of this distribution ($\Delta\theta$) and the angle uncertainties $\delta\theta$ derived from Equation \ref{eq:dtheta} to calculate the dispersion  $\sigma_{\theta}$ using the following relation: $\sigma_{\theta}=\sqrt{\delta\theta^{2}-\Delta\theta^{2}}$. 


In region 4, there appears to be two distinct populations of vectors (see corresponding histogram in Figure \ref{Fig:reg}), so we found the dispersion $\sigma_{\theta}$ for both populations by splitting them between values lesser than and greater than 100$\degree$. We then took the average dispersion for both populations to calculate the magnetic field strength in that specific region.

Using the previously derived values for $n(H_{2})$, $\Delta v$, and $\sigma_{\theta}$ summarized in Table \ref{tab:main}, we calculated the magnetic field strength in each region using Equation \ref{eq:dcf_sim}. The uncertainties in $B_{pos}$ were calculated using:
\begin{equation}
\frac{\delta B_{pos}}{B_{pos}} = \sqrt{\left(\frac{1}{2}\frac{\delta n(H_{2})}{n(H_{2})}\right)^{2} + \left( \frac{\delta \Delta v}{\Delta v}\right)^{2}} \;,
\end{equation}
where $\delta n(H_{2}$) and $\delta \Delta v$ are the uncertainties in volume density and line width, respectively. We calculated the uncertainties in volume density by propagating the uncertainties for the column density. We find magnetic field strengths ranging from $\approx$ 120~$\mu$G to $\approx$ 270~$\mu$G, with fractional uncertainties $\delta B_{pos}/B_{pos}$ ranging from $\approx$ 15\% to $\approx$ 24\%.

\begin{sidewaystable}
	\centering
	\caption{Physical parameters for each of the five regions in L183 (see Figure \ref{Fig:reg}).}
	\scriptsize
	\begin{tabular}{cccccccccccccccccccc}\hline
		Reg &\textit{r} &\textit{L} & $\frac{L}{2r}$  & $f(\frac{L}{D})$  & $N_{H_{2}}$ &$n_{H_{2}}$ &$\Delta v$ & $\rm$\textit{M} &$\sigma_{\theta}$ &$B_{pos}$ &$\lambda_{cor}$ $^{a}$ & $J$ & $|E_{grav}|$ & $E_{th}$ &$E_{mag}$ &$E_{kin}$ & $\frac{E_{mag}}{E_{kin}}$ & $J_{tot}$ \\
		&pc &pc & & &$10^{22}$ $\rm cm^{-2}$ &$10^{5}$ $\rm cm^{-3}$ &$\rm km~s^{-1}$ &$M_{\odot}$ &$\degree$ &$\mu$G & & & $10^{34}$ J & $10^{34}$ J & $10^{35}$ J & $10^{35}$ J & & \\ \hline
		1 &0.016 &0.081 & 2.53 & 1.15876 & $\rm 0.9 \pm 0.2$ &$\rm 1.2 \pm 0.3$ &0.24$\pm$0.03 &0.5$\pm$0.1 &3.7 &206$\pm$33 & 0.11$\pm$0.03 & 2.97 & $7.42 \pm 2.97$ & $2.50 \pm 0.50$ &$\rm 3.2 \pm 1.0$  &$\rm 1.1 \pm 0.5$ &2.8$\pm$0.9 & 0.17 \\
		2 &0.021 &0.077 & 1.83 &0.98055 & $\rm 1.2 \pm 0.3$ &$\rm 1.2 \pm 0.3$ &0.35$\pm$0.03 & 0.9$\pm$0.2 &9.1 &123$\pm$18 & 0.25$\pm$0.07 & 4.40 & $18.3 \pm 8.1$ & $4.16 \pm 0.92$ &$\rm 1.9 \pm 0.6$  &$\rm 3.6 \pm 1.3$ &0.5$\pm$0.2 & 0.34 \\
		3 &0.022 &0.080 & 1.82 & 0.97653 & $\rm 2.8 \pm 1.3$ &$\rm 2.6 \pm 1.2$ &0.28$\pm$0.01& 2.2$\pm$1.0 &4.9 &272$\pm$64 & 0.26$\pm$0.14 & 10.70 & $113 \pm 103$ & $10.6 \pm 4.8$ &$\rm 10.5 \pm 5.0$  &$\rm 6.2 \pm 4.3$ &1.7$\pm$0.8 & 0.68 \\
		4 &0.019 &0.058 & 1.53 &0.89089 & $\rm 1.8 \pm 0.6$ &$\rm 2.0 \pm 0.6$ &0.21$\pm$0.01 & 0.9$\pm$0.3 &6.0 &144$\pm$24 & 0.31$\pm$0.12 & 5.42 & $23.1 \pm 15.4$ & $4.25 \pm 1.42$ &$\rm 1.6 \pm 0.5$  &$\rm 1.6 \pm 0.7$ &1.0$\pm$0.3 & 0.73 \\
		5 &0.029 &0.102 & 1.76 & 0.96039 &$\rm 1.9 \pm 0.8$ &$\rm 1.3 \pm 0.6$ &0.26$\pm$0.01& 2.5$\pm$1.0 &6.6 &135$\pm$28 & 0.36$\pm$0.17 & 9.42 & $113 \pm 90$ & $12.0 \pm 4.8$ &$\rm 5.7 \pm 2.4 $  &$\rm 6.3 \pm 3.5$ &0.9$\pm$0.3 & 0.95 \\ \hline
	\end{tabular}
	\label{tab:main}
	
	a. $\lambda_{cor}$ is the corrected mass-to-flux ratio.
\end{sidewaystable}	


\subsection{Magnetic criticality of the core}\label{sec:critic}

The mass-to-flux ratio $\lambda$ is a unit-less parameter that can be used to quantify the importance of magnetic fields relative to gravity \citep{2004Ap&SS.292..225C}. This parameter can be calculated using the following relation:
\begin{equation}
\lambda=7.6 \times 10^{-21} \frac{N_{H_{2}}}{B_{pos}} \,\, ,
\end{equation}
where $N_{H_{2}}$ is the molecular hydrogen column density in cm$^{-2}$ and $B_{pos}$ is the plane-of-sky amplitude of the magnetic field strength in $\mu$G. When $\lambda < 1$, then the magnetic field is strong enough to restrain the gravitational collapse of the cloud; this is referred as a \enquote{magnetically sub-critical} regime. Alternatively, if $\lambda > 1$, then the magnetic field is insufficient by itself to oppose gravity, and the cloud is instead in a \enquote{magnetically super-critical} state. 

Using our previously derived values for $B_{pos}$ and $N_{H_{2}}$ (see Table~\ref{tab:main}), we calculate the mass-to-flux ratios $\lambda_{obs}$ for each region in L183. We note, however, that these ratios can be overestimated due to geometric biases. For this reason, we followed the same procedure as \citet{2004Ap&SS.292..225C}, and divided our values of $\lambda_{obs}$ by 3 to obtained the corrected mass-to-flux ratios $\lambda_{cor}$ provided in Table~\ref{tab:main}. These results indicate that the L183 cloud as a whole is magnetically sub-critical. 

Based on their survey of molecular lines in starless cores, \citet{Lee2011} initially found L183 to be a potentially contracting core. However, according to their results (see Figure~10 of \citet{Lee2011}), L183 could also be identified as an oscillating core if the classification criteria are slightly relaxed. Our findings that L183 is magnetically sub-critical may therefore help to clarify the dynamical state of this cloud, which was previously unclear using only molecular line data. However, such an expanded analysis would be beyond the scope of this paper.




\subsection{Energy budget of the cloud}\label{sec:ener}

The energy budget of a cloud can be estimated by calculating and comparing its thermal $E_{th}$, kinematic $E_{kin}$, magnetic $E_{mag}$, and gravitational $E_{grav}$ energies. First, we define the following relations:
\begin{equation}
E_{mag} = \frac{B^{2}{V}}{2\mu_{o}} \,\, ,
\label{eq:emag}
\end{equation}
\begin{equation}
E_{th} = \frac{3M\Delta v_{th}^{2}}{2} \,\, ,
\label{eq:eth}
\end{equation}
\begin{equation}
E_{kin} = \frac{3M\Delta v^{2}}{2} \,\, ,
\label{eq:ekin}
\end{equation}
\begin{equation}
\Delta v = \sqrt{(\Delta v_{turb})^{2} + (\Delta v_{th})^{2}} \,\, ,
\label{eq:vel_dis}
\end{equation}
and 
\begin{equation}
    (\Delta v_{th})^{2} = {v^{2}}_{sound} = \frac{k_{B} T_{gas}}{\mu_{free}m_{H}} \, \, ,
    \label{eq:vth}
\end{equation}
where $V$ is the volume, $\mu_{o}$ is the permeability of free space, $\mu_{free}$ is the mean molecular weight of free particles, $\Delta v$ is the total FWHM line width from Equation \ref{eq:vel_dis}, and $\Delta v_{th}$ and $\Delta v_{turb}$ are respectively the contributions of the thermal and turbulent components of this line width. Note that the kinematic energy $E_{kin}$ combines the contribution due to thermal motion of gas particles, as well as the usually stronger energy due to non-thermal supersonic motions from turbulence. The mass $M$ for the region is calculated with $M = n(H_{2}) m_{H} \mu_{H_{2}} \pi r^{2} L$.

In the previous relations, we assume a mean molecular weight $\mu_{free}=2.37$ for a gas mixture of H\,(X = 0.71), He\,(Y = 0.27), and metals (Z = 0.02), but are neglecting the contribution of metals. Furthermore, we adopt the turbulent line widths $\Delta v_{turb}$ from \citet{Lee2001}, and the thermal component $\Delta v_{th}$ was calculated using the measured excitation temperature of 4.6~K from \citet{Pagani2005}.

To find out if the five regions we identified in the L183 cloud are gravitationally bound, we also need to compute their gravitational energies. There is no analytical solution for the gravitational potential of finite uniform cylindrical clouds (Kellogg 1929). Nevertheless, a numerical solution can be expressed as a function $f(L/D)$ of the ratio between the length $L$ and the diameter $D$ of the cylinder \citep{Bastien1979}. The values of this function have been tabulated by \citet{Bastien1983} for typical values of $L/D$ from 0.2 to 10.0. If we define a Jeans number, $J$, one can show that
\begin{equation}
    J_{cyl} = \frac{|E_{grav}|}{E_{th}}=\frac{G \, m}{\frac{k_{B}T}{\mu_{free}m_{H}}} \; f\left( \frac{L}{D} \right) \, \, ,
    \label{eq:Jcy1}
\end{equation}
\begin{equation}
    J_{cyl} = 3\frac{m}{m_{c}} \; f\left(\frac{L}{D}\right) \, \, ,
    \label{eq:Jcy2}
\end{equation}
where $m = M/L$ is the linear mass of the cloud. 

The critical linear mass for an infinite cylinder is given by:
\begin{equation}
    m_{c} = \left(\frac{M}{L}\right)_{c} = \frac{3 k_{B} T}{b \, G \, \mu_{free} m_{H}} \, \, .
    \label{eq:mc}
\end{equation}
The constant $b$ depends on the density distribution (i.e., $b = 1$ for uniform density cylinders \citep{McCrea1957}, and $b=3/2$ for equilibrium cylinders \citep{Ostriker1964}). When $m > m_{c}$, the filament (cylinder) collapses along its axis. Otherwise, if $m < m_{c}$, the infinite cloud will not collapse, even by increasing the external pressure. The values of $f(L/D)$ for our five regions were determined by a linear interpolation of $log[f(L/D)]$ between the known values in Table~1 from \citet{Bastien1983}. We used Equation \ref{eq:mc} with $b=1$ for uniform density cylinders to get $m_{c} = 7.5 M_{\odot}/pc$ assuming a temperature $T = 4.6\,K$. Combining Equations \ref{eq:eth} and \ref{eq:Jcy1}, we obtain the gravitational energy $E_{grav}$ of the cylinders, which can be found in Table~\ref{tab:main}. The value of $E_{grav}$ can be found using equation below.

\begin{equation}
    E_{grav} = -\frac{9}{4}\frac{G M^{2}}{b L} f(L/D)
    \label{eq:Egrav}
\end{equation}

The derived values of the Jeans numbers (=$1/\alpha$ in many other works) are all $>\approx 3.0$, and are larger than the critical Jeans numbers for cylinders with $1.5<L/D<2.5$ which is $\approx 0.8$ \citep{Bastien1983}. This means that all five regions are gravitationally bound if we consider only gravity and thermal pressure. Moreover, they can accommodate other forces that tend to counter gravity, such as turbulence and magnetic fields. 

To take into account other forces, we computed the following quantity:
\begin{equation}
    J_{tot} = \frac{E_{grav}}{E_{kin} + E_{mag}}.
    \label{eq:Jtot}
\end{equation}
We see that all the regions in L183, except maybe region 5 (see Figure \ref{Fig:reg}), will not be bound when thermal and non-thermal motions, in addition to magnetic fields, are taken into account. However, we have to be careful since the contribution of a magnetic field depends on its configuration. For example, a toroidal field will constrain the gas on the axis of the filament, and a poloidal field aligned with the axis will increase motions along that axis. In any case, as shown with hydrodynamics calculations, gravity produces significant motions of material along the axis of the filament. After some time, these motions increase the line mass and will make it possible to get sections of the filament to become gravitationally unstable, even if they were originally stable. However, a truly infinite cylindrical cloud will
not produce motions along its axis, unless it has density perturbations
along its axis.

\subsection{Comparison with previous polarization measurements in L183}\label{sec:comp}

\begin{figure}
	\centering
	\resizebox{8.5cm}{5.8cm}{\includegraphics{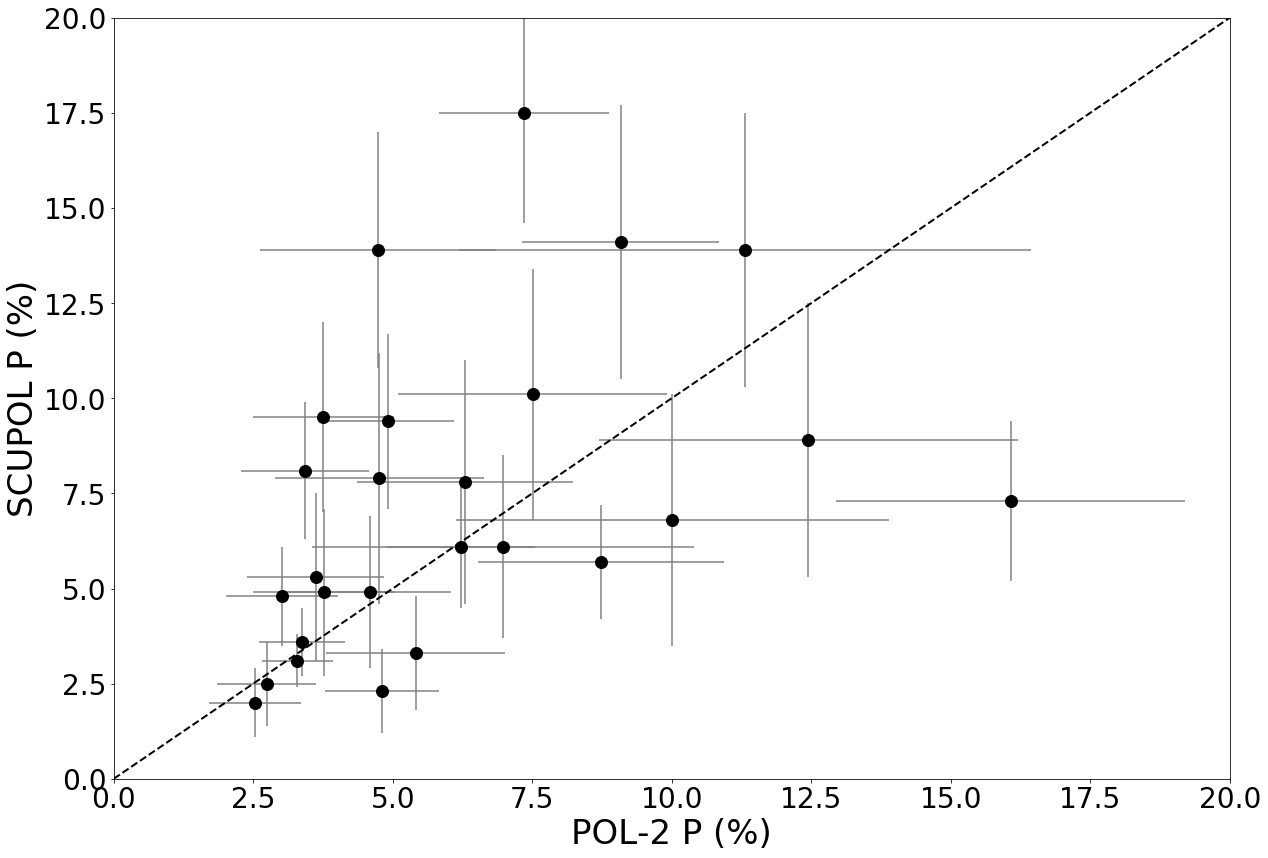}}
	\resizebox{8.5cm}{5.8cm}{\includegraphics{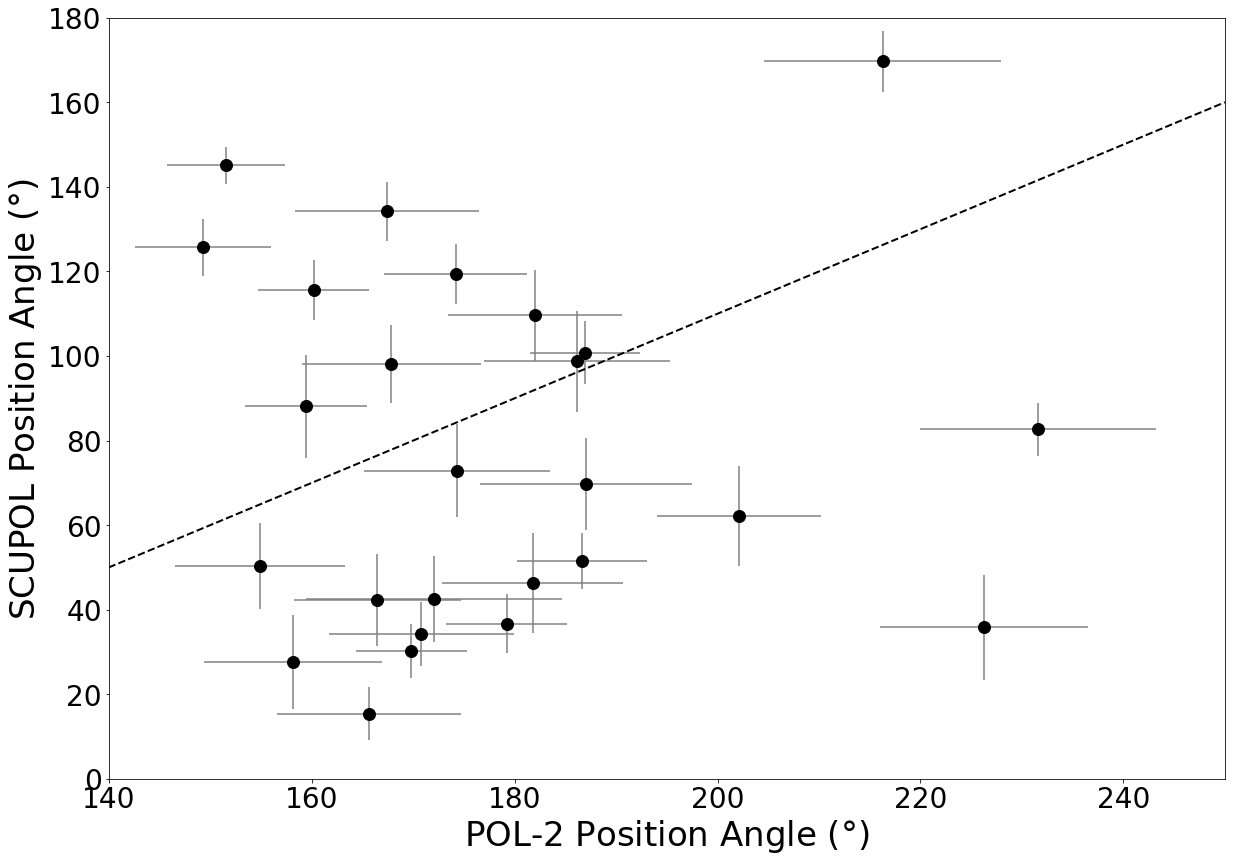}}
	\caption{The upper panel shows a comparison between SCUPOL \citep{2009ApJS..182..143M} and SCUBA-2/POL-2 (this work) polarization percentages measured toward L183. Similarly, the lower panel shows a comparison of polarization angles. In both plots, the dashed line shows the corresponding one-to-one relation.}
	\label{Fig:scupol}
\end{figure}

\citet{2009ApJS..182..143M} reanalyzed the polarization data previously obtained towards L183 using SCUBA-2/POL-2's predecessor, SCUPOL \citep{2000ApJ...537L.135W,2004ApJ...600..279C}. They reported 26 data points with P/$\delta P>$2 towards the main core of L183, while our observations yield 236 such data points in the whole filamentary cloud structure (including the main core). The two data sets are spatially coincident to within $\sim 6\farcs$. In Figure~\ref{Fig:scupol}, the comparisons between SCUPOL and POL-2 polarization percentages (upper panel) and position angles (lower panel) for L183 are plotted. The degrees of polarization in both samples agree within a 3$\sigma$ range. We added 180$\degree$ to the SCUBA-2/POL-2 position angles that are less than 90$\degree$ to try and find a better relationship between SCUPOL and POL-2 detections. Our data achieve a higher SNR and sensitivity both in the region studied by \citet{2009ApJS..182..143M} and \citet{2004ApJ...600..279C} and in the rest of the large-scale filament, and so the observed differences are likely due to the increased sensitivity of SCUBA-2/POL-2 relative to SCUPOL.

We also compared our results to the magnetic field strength estimates from the SCUPOL data by limiting our DCF analysis to only the region of L183 analyzed by \citet{2004ApJ...600..279C}. Following their method, we treated the region as a sphere and used the velocity dispersion value of 0.22 $\rm km~s^{-1}$ from \citet{2002ApJ...572..238C}. We obtained $B_{pos}=105\pm24\mu G$, consistent with $\sim\,80\mu G$ from \citet{2004ApJ...600..279C}.

\section{Summary} \label{sec:summary}
\begin{enumerate}
	\item We presented the deepest 850 $\mu$m continuum Stokes~$I$, $Q$, and $U$ observations to date of the starless cloud L183.  The Stokes \textit{I} map shows an elongated filamentary structure containing two distinct dense cores, as well as several additional, less dense condensations.
	\item We compared the magnetic field morphology derived from POL-2 data to that of optical and near infrared data. We found that, while the large-scale field in the extended cloud run in an east-west direction, the magnetic fields in the cores are predominantly oriented north-south along the direction of the filament's elongation.
	\item The L183 filamentary structure separates into five sub-regions for which we performed individual analysis of the polarization degree, as well as of the orientation and strength of the magnetic field. Out of these five regions, region 3, which contains the main core of L183, is found to be the densest ($n_{H_2} = 2.6\times10^{5}$~cm$^{-3}$) with the strongest magnetic field ($B=272$~$\mu$G). All other relatively diffuse regions have similar density and field strengths. 
	\item We estimated the gas column density and the dust temperature of the mapped region by supplementing our 850~$\mu$m data with \textit{Herschel} SPIRE/PACS continuum observations. The average values of column density and temperature in the filament are $\rm \sim1.5\times10^{22}~cm^{-2}$ and $\sim$10~K, respectively. 
	\item The magnetic field strength in each of the mapped regions ranges from $\sim120\pm18~\mu G$ to $\sim270\pm64~\mu G$. With our derived field strength and column density, we calculated the criticality parameter $\lambda_{cor}$ in these five regions, and found values ranging from 0.1 to 0.4. These results suggest that the L183 is magnetically sub-critical everywhere, except for region 5 which could be gravitationally bound because of its lower magnetic energy and somewhat larger mass. 

\end{enumerate}

\bigskip
\bigskip
We thank the anonymous referee for an encouraging and constructive report, which helped in improving the final manuscript. J.K. is supported by Santa Clara University. A.S. and B-G.A. are supported by NSF Grant-1715876. This research was conducted in part at the SOFIA Science Center, which is operated by the Universities Space Research Association under contract NNA17BF53C with the National Aeronautics and Space Administration. C.W.L. is supported by the Basic Science Research Program through the National Research Foundation of Korea (NRF) funded by the Ministry of Education, Science and Technology (NRF-2019R1A2C1010851). J.E.V. acknowledges financial support from NASA through award \# SOF 05-0038 issued by USRA. The JCMT is operated by the East Asian Observatory on behalf of National Astronomical Observatory of Japan; Academia Sinica Institute of Astronomy and Astrophysics; the Korea Astronomy and Space Science Institute; the Operation, Maintenance and Upgrading Fund for Astronomical Telescopes and Facility Instruments, budgeted from the Ministry of Finance of China. SCUBA-2 and POL-2 were built through grants from the Canada Foundation for Innovation. This research used the facilities of the Canadian Astronomy Data Centre operated by the National Research Council of Canada with the support of the Canadian Space Agency. 

This research made use of Astropy,\footnote{http://www.astropy.org} a community-developed core Python package for Astronomy \citep{2013A&A...558A..33A, astropy:2018}. The Starlink software \citep{2014ASPC..485..391C} is currently supported by the East Asian Observatory.

%

\facility{JCMT (SCUBA-2, POL-2)}


\software{Starlink \citep{2014ASPC..485..391C}, Astropy \citep{2013A&A...558A..33A}.}



\bibliographystyle{aasjournal}



\end{document}